\documentclass[twocolumn,aps,showpacs,multicol,amsmath,amssymb]{revtex4-1}
\usepackage{amssymb}
\usepackage{amsmath}
\usepackage{graphicx}
\usepackage{epstopdf}
\usepackage{bm}% bold math
\usepackage{gensymb}
\usepackage{soul}
\usepackage{sidecap}
\usepackage[normalem]{ulem}
\usepackage{float}
\usepackage[caption = false]{subfig}
\usepackage{subfig}
\usepackage{enumerate}
\usepackage{multirow}
\usepackage{tabularx}
\usepackage{array}
\usepackage{url}
\usepackage{slantsc}
\usepackage{lmodern}
\newcommand\redout{\bgroup\markoverwith{\textcolor{red}{\rule[.5ex]{2pt}{0.4pt}}}\ULon}

\usepackage[colorlinks=true,citecolor=blue]{hyperref}
\hypersetup{colorlinks=true,citecolor=blue,linkcolor=red,urlcolor=blue}

\usepackage{hyperref,cleveref}

\newcommand{\be}{\begin{equation}}
\newcommand{\ee}{\end{equation}}
\newcommand{\bk}{{{\bf{k}}}}
\newcommand{\bq}{{{\bf{q}}}}

\newcommand{\br}{{{\bf{r}}}}

\newcommand{\bg}{{{\bf{g}}}}
\newcommand{\bG}{{{\bf{G}}}}

\newcommand{\htau}{{\hat{\tau}}}

\newcommand{\bea}{\begin{eqnarray}}
\newcommand{\eea}{\end{eqnarray}}

\newcommand{\bd}{\begin{displaymath}}
\newcommand{\ed}{\end{displaymath}}
\newcommand{\ba}{\begin{array}}
\newcommand{\ea}{\end{array}}
\newcommand{\bi}{\begin{itemize}}
\newcommand{\ei}{\end{itemize}}
\newcommand{\bc}{\begin{center}}
\newcommand{\ec}{\end{center}}
\newcommand{\bfl}{\begin{flushleft}}
\newcommand{\efl}{\end{flushleft}}
\newcommand{\bfr}{\begin{flushright}}
\newcommand{\efr}{\end{flushright}}
\newcommand{\no}{\nonumber}

\newcommand{\mi}{\rm i}
\newcommand{\inv}{^{\raisebox{.2ex}{$\scriptscriptstyle-1$}}}
\def\br{{\bf r}}
\def\bk{{\bf k}} \def\bq{{\bf q}}  
\def\bg{{\bf g}}  \def\bd{{\bf d}}

\def\6{\partial}

\def\={\!\!\!&=&\!\!\!}
\def\+{\!\!\!&&\!\!\!+~}
\def\-{\!\!\!&&\!\!\!-~}

\graphicspath{{.}{./Figs/}}
\begin{document}
\title{ Drumhead  surface states and their signatures in quasiparticle scattering interference}
 \author{Mehdi Biderang$^{1,2}$}
 \thanks{These two authors contributed equally to this work.}
 \author{Andreas Leonhardt$^{3}$}
 \thanks{These two authors contributed equally to this work.}
 \author{Nimisha Raghuvanshi$^{1}$}
 \author{Andreas  P. Schnyder$^{3}$}
 \author{Alireza Akbari$^{1,4,5}$}\email{alireza@apctp.org}
\affiliation{$^1$Asia Pacific Center for Theoretical Physics, Pohang, Gyeongbuk 790-784, Korea}
\affiliation{$^2$Department of Physics, University of Isfahan, Hezar Jerib, 81746-73441, Isfahan, Iran}
\affiliation{$^3$Max-Planck-Institute for Solid State Reserach, Heisenbergstra{\ss}e 1, D-70569 Stuttgart, Germany}
\affiliation{$^4$Department of Physics, POSTECH, Pohang, Gyeongbuk 790-784, Korea}
\affiliation{$^5$Max Planck POSTECH Center for Complex Phase Materials, POSTECH, Pohang 790-784, Korea}
%%%%%%%%%%%%%%%%%%%%%%%%%%%%%%%%%%%%%%%%%%%%%%%%%%%%%%%%%%%%%%%%%%%%%%%%%%%
\date{\today}
%
%%%%%%%%%%%%%%%%%%%%%%%%%%%%%%%%%%%%%%%%%%%%%%%%%%%%%%%%%%%%%%%%%%%%%%%%%%%
\begin{abstract}
 We consider a two-orbital tight-binding model  defined on a layered three-dimensional hexagonal lattice to investigate the
properties of topological nodal lines and their associated drumhead surface states.
We examine these surface states in centrosymmetric
systems,
where the bulk nodal lines are of Dirac type (i.e., four-fold degenerate), 
as well as in non-centrosymmetric systems with strong Rashba and/or Dresselhaus   spin-orbit coupling,
 where the bulk nodal lines are of Weyl type (i.e.,  two-fold  degenerate).
We find that in non-centrosymmetric systems the nodal lines and their corresponding drumhead surface states are fully spin polarized due to 
 spin-orbit coupling.   We show that unique signatures of the topologically nontrivial drumhead surface states can be measured by means of 
 quasiparticle scattering interference, which we compute for both Dirac and Weyl nodal line semimetals.
 \end{abstract}
%%%%%%%%%%%%%%%%%%%%%%%%%%%%%%%%%%%%%%%%%%%%%%%%%%%%%%%%%%%%%%%%%%%%%%%%%%%
%
\maketitle
%
%%%%%%%%%%%%%%%%%%%%%%%%%%%%%%%%%%%%%%%%%%%%%%%%%%%%%%%%%%%%%%%%%%%%%%%%%%%
%%%%%%%%%%%%%%%%%%%%%%%%%%%%%%%%%%%%%%%%%%%%%%%%%%%%%%%%%%%%%%%%%%%%%%%%%%%
\section{Introduction}
 The discovery of time-reversal invariant topological insulating
materials with strong spin-orbit coupling (SOC) triggered an
enormous research interest in decoding the topological phases of
matter \cite{hasan_2010,qi_2011,chiu_2016}.  One of the most
important characteristics of topological materials is the
existence of symmetry-protected metallic modes at the surface of
the material, while the bulk shows insulating or semi-metallic
behaviour.  This property originates from the topologically
nontrivial ordering of bulk wave functions and is characterized by
a band inversion involving switching of bands of opposite parity
around the Fermi level \cite{neupane_2013,hasan_2015}.  The recent
theoretical prediction and experimental observation of topological
nodal-line semimetals
\cite{bian_2016,bian_a_2016,burkov_2016,chan_2016,chang_2016,fang_2016,schoop_2016,neupane_2016,wu_2016,hu_2016,okamoto_2016,takane_2016,yu_2017,kobayashi_2017},
Weyl semimetals
\cite{wan_2011,xu_2011,yang_2011,singh_2012,witczak_2012,huang_2015,xu_2015,lu_2015,lv_2015,xu_a_2015,zheng_2016},
and Dirac semimetals
\cite{young_2012,wang_2012,wang_2013,liu_2014,yang_2014,neupane_2014,he_2014,liu_a_2014,gibson_2015,pariari_2015,yu_2015,yamakage_2016,liu_2016}
has redirected the research interest from insulating topological
materials to topological semimetals in the field of quantum
condensed matter physics.

The defining feature of topological semimetals is the crossing of
the conduction and the valence bands in the Brillouin zone (BZ)
near the Fermi energy.  These crossings are 
%non-accidental and
topologically protected against gap opening by any perturbation
that preserves a certain symmetry group.  These bulk band
degeneracies can occur at discrete nodal-points or continuous
nodal-lines resulting in zero-dimensional bulk nodal-points
(Weyl/Dirac semimetals) or one-dimensional bulk nodal-lines
(topological nodal-line semimetals). 
Weyl semimetals can be realized when either time reversal or
inversion symmetry is broken, while the stability of Dirac nodal
lines is guaranteed  by time-reversal symmetry combined with
inversion symmetry  or a crystal symmetry, such as rotation or
reflection.  Therefore, a nodal line can either evolve from a
crossing of two bands or  two doubly degenerate bands, thus
exhibiting Weyl character or Dirac character.  In the following we
call these two cases Weyl nodal-line semimetal (WNLS) and Dirac
nodal-line semimetal (DNLS).  In the Dirac and Weyl topological
semimetallic states, the low-energy excitations are linearly
dispersing Dirac and Weyl fermions.  In WNLSs and DNLSs, on the
other hand, the Weyl/Dirac fermions are linearly dispersing in
only two momentum directions, while they are non-dispersive in the
third direction, i.e., along the nodal line.
% Dirac fermions with linear dispersions along all momentum
% directions can be viewed as two copies of distinct Weyl
% fermions. 

%%%%%%%%%%%%%%%%%%%%%%%%% Figure %%%%%%%%%%%%%%%%%%%%%%%%%%%%%%%%%
\begin{figure}[t]
    \begin{center}
    % TODO: find a matching crystal system
    % TODO: drop B-atoms? Keep them to break inversion symmetry?
    \includegraphics[width= 0.95\linewidth]{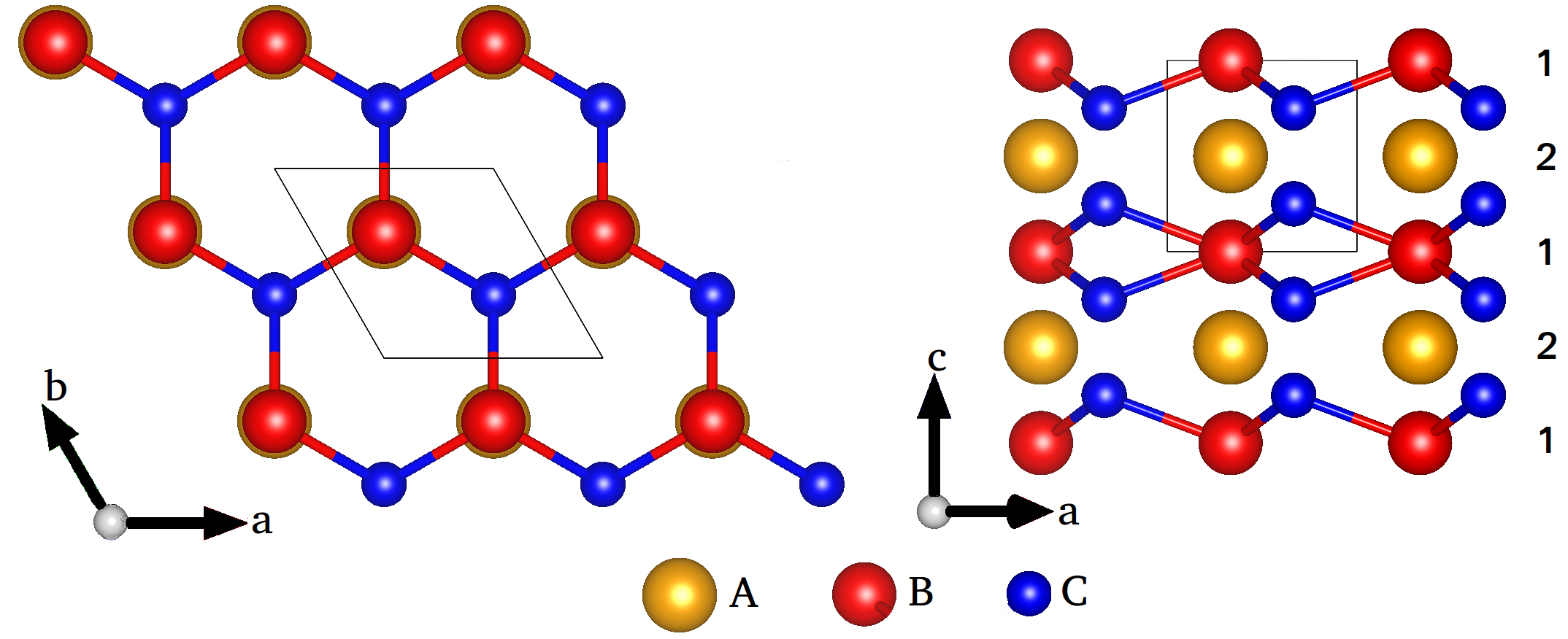}
    \end{center}     \vspace{-.4cm}
      \caption{ Crystal structure of the three-dimensional layered
                system as viewed from the [001] (left panel) and
                the [010] (right panel) directions of the
                hexagonal lattice.  Layer $1$ (with B\&C atoms)
                intercalates between the adjacent $2$ layers such
                that A atoms are aligned with the B atoms in the
                $z$-direction.
            }
    % TODO: adjust text
       \label{Fig1}%
\end{figure}
%%%%%%%%%%%%%%%%%%%%%%%%%%%%%%%%%%%%%%%%%%%%%%%%%%%%%%%%%%%%%%%%%

%%%%%%%%%%%%%%%%%%%%%%%%%Figure%%%%%%%%%%%%%%%%%%%%%%%%%%%%%%%%%%
\begin{figure*}[t]
 \begin{center}
     \includegraphics[width=0.85\linewidth]{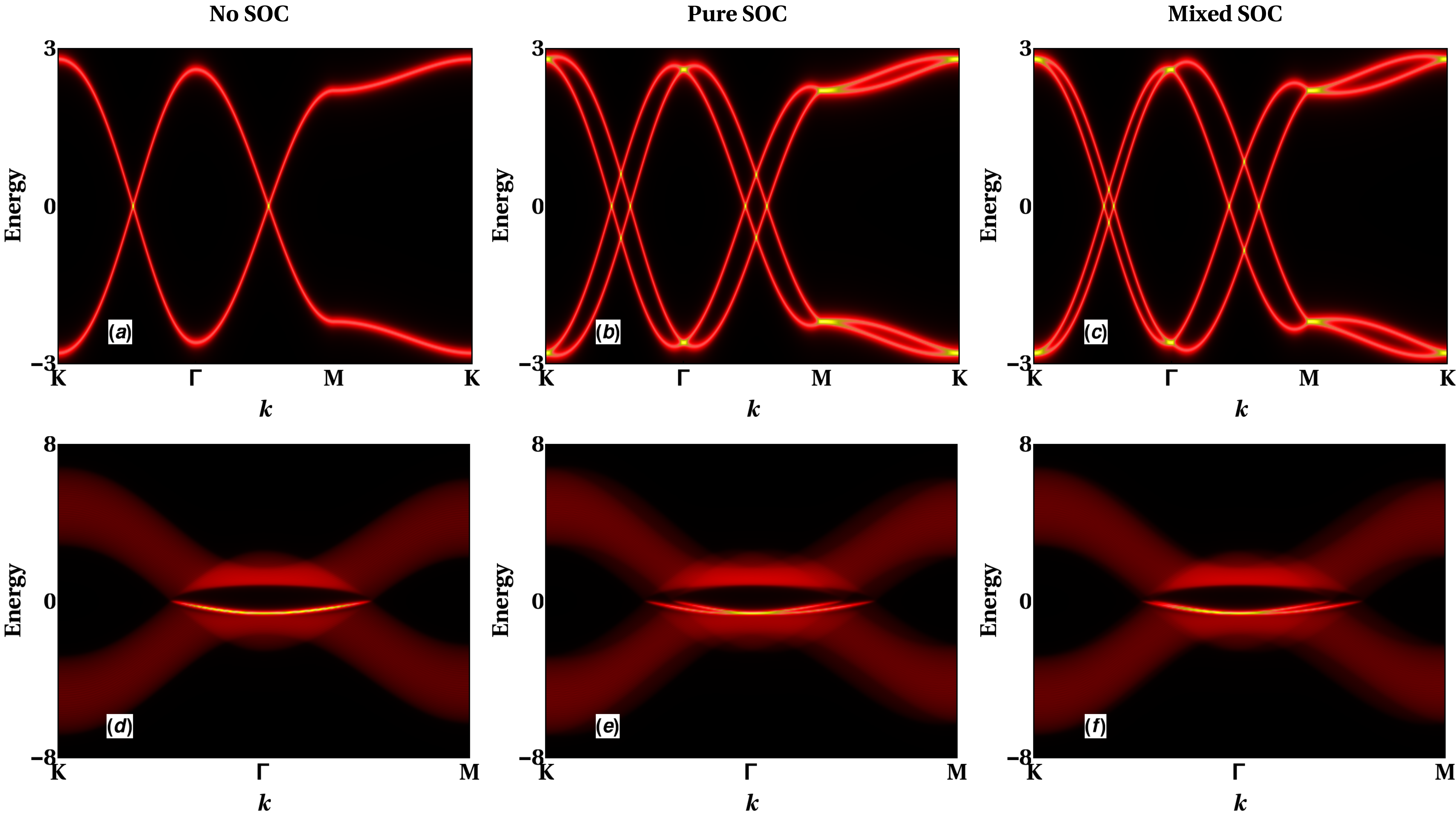}
\end{center}
\vspace{-0.5cm}
      \caption{
     Upper panel:      Bulk energy dispersion  at $k_z$=0 along  high symmetry
        lines in the BZ for (a) the Dirac nodal line semimetal, 
        (b) the Weyl nodal line semimetal with pure Rashba 
            ($\alpha=0.2, \beta=0.0$) or Dresselhaus 
            ($\alpha=0.0, \beta=0.2$) SOCs, and 
        (c) the Weyl nodal line semimetal with mixed Rashba and 
            Dresselhaus ($\alpha=\beta=0.1$) SOC. Here and in the 
            rest of the paper, energies are measured in units of 
            $t_\bot$.% 
   Lower panel: Energy dispersion of the surface density of states as a
    function of energy and momentum along  high symmetry lines in 
    the BZ for 
    (d)  the  Dirac nodal line semimetal, 
    (e) the  Weyl nodal
        line semimetal (WNLS) with pure Rashba SOC (or pure
        Dresselhaus SOC), and 
    (f) the WNLS with mixed Rashba and Dresselhaus SOC.
      }
    \label{Fig2}%
\end{figure*}
%%%%%%%%%%%%%%%%%%%%%%%%%%%%%%%%%%%%%%%%%%%%%%%%%%%%%%%%%%%%%%%%%%
%
%%%%%%%%%%%%%%%%%%%%%%%%%%%%%%%%%%%%%%%%%%%%%%%%%%%%%%%%%%%%%%%%%%

A nodal line which closes inside the BZ forming a nodal loop and
carrying a topological charge, is associated with drumhead-like
surface states
\cite{burkov_2011,kopnin_2011,zhao_2013,matsuura_2013,chiu_2016}.
These surface states are bounded by the  projection of the nodal
ring onto the surface, a region looking like the head of an open
drum and hence are called drumhead surface states.  
Thus, the intricate interplay between symmetry and topology of the
electronic wave functions gives rise to the associated
two-dimensional drumhead-like surface states in topological
nodal-line semimetals.  Topological nodal-line semimetals
exhibiting line-node bulk states and associated drumhead surface
states present a significant expansion of topological materials
beyond topological insulators and nodal point semimetals.  They
provide new opportunities to explore the exotic physics of
topological materials.  The novel properties of topological
nodal-line semimetals, like drumhead surface states
\cite{weng_2015,yu_2015,kim_2015,volovik_2013}, have opened new
possibilities in  high-$T_C$ superconductivity
\cite{kopnin_2011,tang_2014,volovik_2015,heikkila_2016,wang_2016,
        chang_2016,pang_2016}
and magnetic order \cite{magda_2014}, unique Landau energy levels
\cite{rhim_2015}, collective modes from the novel nodal line
structure \cite{yan_2016}, and  anomalies in electromagnetic and
transport response
\cite{weng_a_2015,chiu_2014,yang_a_2014,burkov_2011,phillips_2014,
        huh_2016,matusiak_2017,emmanouilidou_2017,mukherjee_2017,
        rui_zhao_schnyder_arXiv_17}.

% TODO: rewrite such as to end up with a model for 
% CaAg(P/As)
Many nodal line materials have been experimentally realized
recently, like PbTaSe$_2$ \cite{bian_2016}, Ca$_3$P$_2$
\cite{xie_2015,chan_2016} CaAgP and  CaAgAs
\cite{yamakage_2016,okamoto_2016,takane_2017,nayak_felser_JPCM_18,
        wang_chang_PRB_17}.
Other candidate materials include TlTaSe$_2$ \cite{bian_a_2016},
fcc alkaline earth metals \cite{hirayama_2017}, hyperhoneycomb
lattices \cite{mullen_2015} and the CaP$_3$ family \cite{xu_2017}.
Noncentrosymmetric PbTaSe$_2$ \cite{chang_2016} and
TlTaSe$_2$ \cite{bian_a_2016}  have reflection symmetry at the Ta
atomic plane but lack inversion symmetry, and hence exhibit Weyl
rings with two-fold degeneracy, while
fcc alkaline earth metals, hyperhoneycomb lattices and Ca$_3$P$_2$
are protected by both 
inversion and time-reversal symmetry, and thus display Dirac rings
which are four-fold degenerate.  The noncentrosymmetric hexagonal
ternary pnictides, CaAgP, CaAgAs,  and SrPtAs are particularly
promising candidates, since they can be grown in single crystal
form~\cite{takane_2017,nayak_felser_JPCM_18,wang_chang_PRB_17,
Fischer:2015aa,Biswas:2013aa,Goryo:2012aa,Akbari:2014aa,
Fischer:2014aa}.

In this paper, we  consider a two-orbital tight-binding model to
study the topological properties of a three-dimensional layered
hexagonal semimetal with both time-reversal and reflection
symmetries.  We study the effects of the presence and absence of
inversion symmetry on the electronic band structure and topology
of the surface states.  In particular, we examine how Rashba and
Dresselhaus spin-orbit couplings modify the properties of the
nodal line and the drumhead surface states. 

Furthermore, we compute the quasiparticle interference (QPI)
patterns for both Dirac and Weyl nodal-line semi-metals and
identify the signature of the drumhead surface states in these QPI
patterns.

%%%%%%%%%%%%%%%%%%%%%%%%%%%%%%%%%%%%%%%%%%%%%%%%%%%%%%%%%%%%%%%%%%
\section{Model} 
%
%{\AL
%We study the topology of the band structure in a nodal line
%semimetal by means of a generic two band tight binding model. Our
%system consists of consecutive two-dimensional hexagonal layers,
%aligned in the 001 direction.  Each layer is mapped onto itself
%under reflection symmetry, with the orbitals of one layer
%being symmetric under $R_z$, while the other is
%antisymmetric.  
%In the experimentally realized DNLSM CaAgP for example, these are
%formed by the $3p_z$ orbitals of one of the P atoms and the
%symmetric combination of the three $3s$ orbitals of Ag, which are
%then centered around a common axis.
%%
%These two orbitals form particle- and hole-like bands, which cross
%near the Fermi energy, generating a line-like degeneracy. 
%The different $R_z$ symmetry eigenvalues prevent hybridization,
%keeping the line nodes in semimetals stable against gap opening.
% 
%We consider two different scenarios with and without antisymmetric
%spin-orbit coupling (SOC),  describing Weyl and Dirac nodal line
%semimetals, respectively.  
% }

In this section, we survey the topology of the band structure in a
nodal line semimetal by means of a tight-binding model with two
orbitals.   
These two orbitals form   particle- and hole-like bands, which
cross each other, generating a line-like degeneracy that is
topologically protected.  We consider two different scenarios with
and without antisymmetric spin-orbit coupling (SOC),  describing
Weyl and Dirac nodal line semimetals, respectively.  
Our system consists of consecutive two-dimensional hexagonal
layers of $s$ and $p_z$ orbitals, which are invariant under
reflection symmetry ($R_z$) along the $z$-direction.
Thus, the bands that cross near the Fermi energy have opposite
$R_z$ symmetry eigenvalues, which prevents hybridization,  keeping
the line nodes in semimetals stable against gap opening. 
Fig.~\ref{Fig1} shows the crystal structure for this kind of
system with lattice parameters $a$ and $c$.  The orbitals of the
tight binding model are located at sites A and B, while the C-atoms
break inversion symmetry.
Our model can be assumed as a general description for the
realization of topological drumhead surface states in a DNLSs like
CaAgP and CaAgAs, as well as, WNLSs such as the ternary
chalcogenides TlTaSe$_2$ and PbTaSe$_2$.
In the latter ones, the nodal line lies in the $k_z = \pm \pi$
plane and closes around the $H$ point of the hexagonal BZ.
The relevant orbitals there are $d$ and $p_{x,y}$, and the
different symmetry eigenvalues come from the factor 
$\rm e^{\mi k_z/2}$ for the atom located $c/2$ above the origin.

\subsection{Dirac nodal line semimetal}
We begin our analysis by discussing the Dirac nodal line
semimetals which are invariant under both time-reversal and
spatial inversion symmetry.  This implies that both external
magnetic field and antisymmetric SOC are absent.  Without
antisymmetric SOC the particle- and hole-like bands  are
spin-degenerate.  Time-reversal, inversion, and $SU(2)$
spin-rotation symmetry together with a non-trivial band topology
lead to a protected four-fold degenerate band crossing, which
forms a nodal ring~\cite{chan_2016}.  Based on this picture, we
consider a spinless two-orbital tight-binding model with the
Hamiltonian
%
%================================================================
%===================== Formula ==================================
%================================================================
\begin{align} \label{H_r}
\begin{aligned}
{\cal H}_0=
&
\mu \sum_{i,\gamma\gamma'} 
 \htau_z^{\gamma\gamma'} c^{\dag}_{i,\gamma}c^{}_{i,\gamma'}
\\
&
-\sum_{\langle ij \rangle,\gamma\gamma'}
\Big[
(
 t_{ij}
 \htau_z^{\gamma\gamma'}  
-  
{\mi}
 t^{\prime}_{ij}
  \htau_y^{\gamma\gamma'}
)
c^{\dag}_{i,\gamma}c^{}_{j,\gamma'}
+ h.c.  
  \Big],
\end{aligned}
\end{align}
%===========================================================
where $c^{\dag}_{i,\gamma}$ ($c_{i,\gamma}$) is the creation
(annihilation) operator for an electron on site $i$ and orbital
$\gamma~(=p,d)$.  Here, $t_{ij}$ and $t^{\prime}_{ij}$ denote
intra- and inter-orbital hopping amplitudes. 
Intra-orbital hopping is both, out of plane between first neighbours
and in-plane between second neighbours, parametrized by 
$t_\perp$ and $t_\parallel$.
Inter-orbital hopping is in between the layers and parametrized
by $t^\prime$.

%which contain both
%intra-layer ($t_{\parallel}$, $t^{\prime}_{\parallel}$) and
%inter-layer ($t_\perp$, $t^{\prime}_\perp$) terms.  
Moreover,
$\htau_i$ ($i=x,y,z$) are the Pauli matrices in orbital space and
$\mu$ is an on-site energy, with opposite sign for $p$ and $d$
orbitals.  For the numerical calculations, we set the values of
the physical parameters to
$(\mu,t_{\parallel},t_{\perp}, t^{\prime})=(3,0.6,1,0.5)$. 
In momentum space,
the Hamiltonian in Eq.~(\ref{H_r}) is given by
%===========================================================
%===================== Formula =============================
%===========================================================
\begin{subequations} \label{H_k}
\begin{align}   
\begin{aligned}
{\cal H}_0=
\sum_{\bk,\gamma\gamma'} 
\Big[
 \varepsilon_1(\bk)~ \htau_z^{\gamma\gamma'} + \varepsilon_2(\bk)~ \htau_y^{\gamma\gamma'}
\Big ] c^{\dag}_{\bk,\gamma}c_{\bk,\gamma'},
\end{aligned}
\end{align}
%===========================================================
where
%===========================================================
%===================== Formula =============================
%===========================================================
\begin{align} \label{epsi_1}
%\no
\begin{aligned}
\varepsilon_1(\bk)=
&
\;\mu-2t_{\perp}\cos(k_zc)
\\
&
-2 t_{\parallel} 
\Big[
2\cos(\frac{3k_xa}{2}) \cos(\frac{\sqrt{3}k_ya}{2})
+\cos(\sqrt{3}k_ya) \Big]
\end{aligned}
\end{align}
and
%===========================================================
%
\begin{align} \label{epsi_2}
%\no
\begin{aligned}
\varepsilon_2(\bk)=
&
-2t^{\prime}\sin(k_zc/2)
%\\
%&
%+2 t^{\prime}_{\parallel} 
%\Big[
%2\cos(\frac{k_xa}{2}) \sin(\frac{\sqrt{3}k_ya}{2})+\sin(k_xa) 
%\Big]
.
\end{aligned}
\end{align}
\end{subequations}
%===========================================================
%
Figure~\ref{Fig2}(a) shows the energy dispersion of
Hamiltonian~\eqref{H_k} along the K$\Gamma$MK-path within the
$k_z=0$ plane of the BZ, which is invariant under $R_z$. It can be
seen that around the $\Gamma$ point, two two-fold degenerate hole-
and electron-pockets cross each other, generating a four-fold
degenerate Dirac nodal ring. Figure~\ref{Fig3}(a) depicts the
whole Dirac nodal ring at the $k_z=0$ plane in the hexagonal BZ.  

%%%%%%%%%%%%%%%%%%%%%%%%%Figure%%%%%%%%%%%%%%%%%%%%%%%%%%%%%%
\begin{figure}[]
    \begin{center}
 \hspace{-0.1cm}    
  \includegraphics[width=0.323 \linewidth]{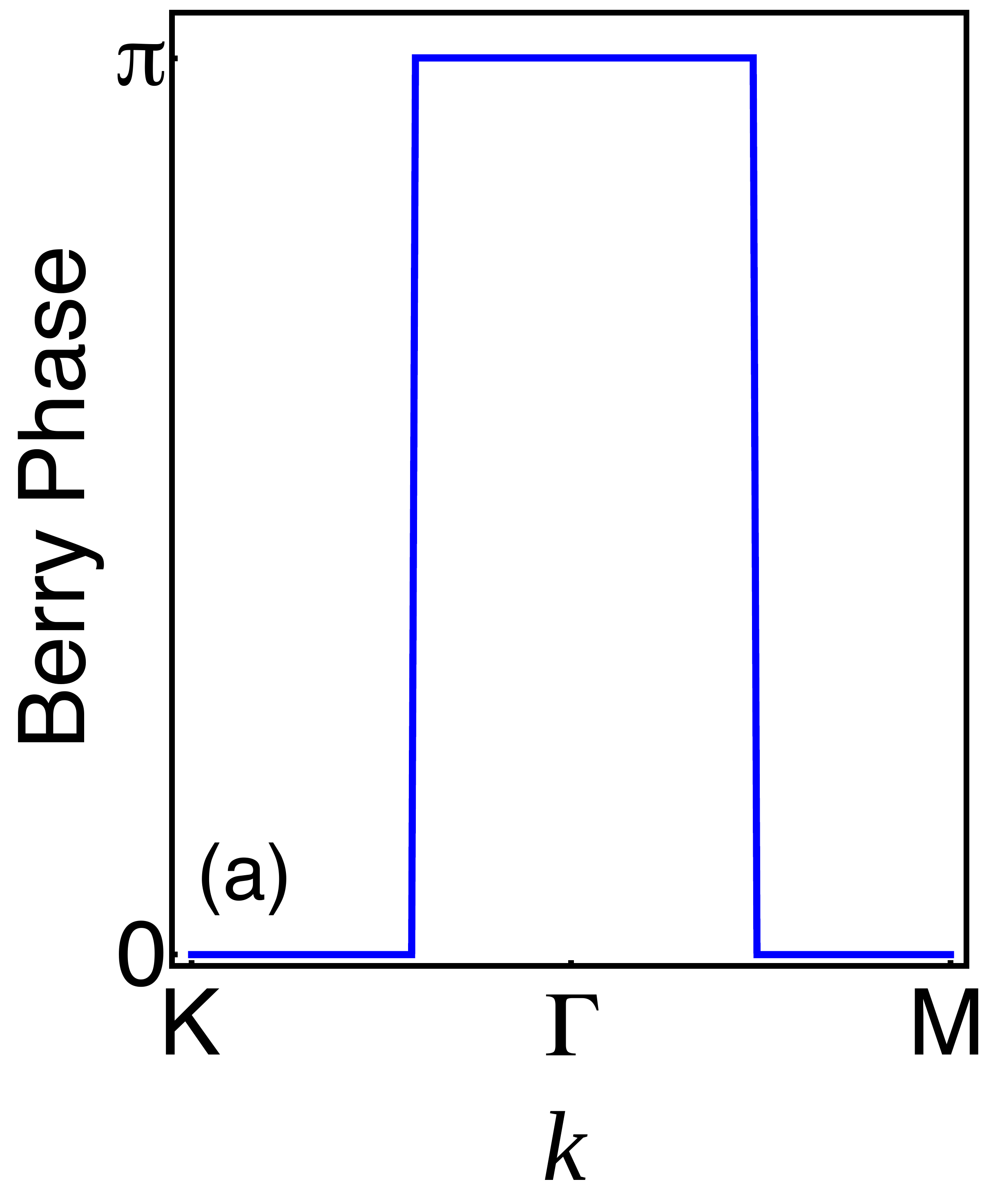}
   \hspace{0.1cm}    
  \includegraphics[width=0.635\linewidth]{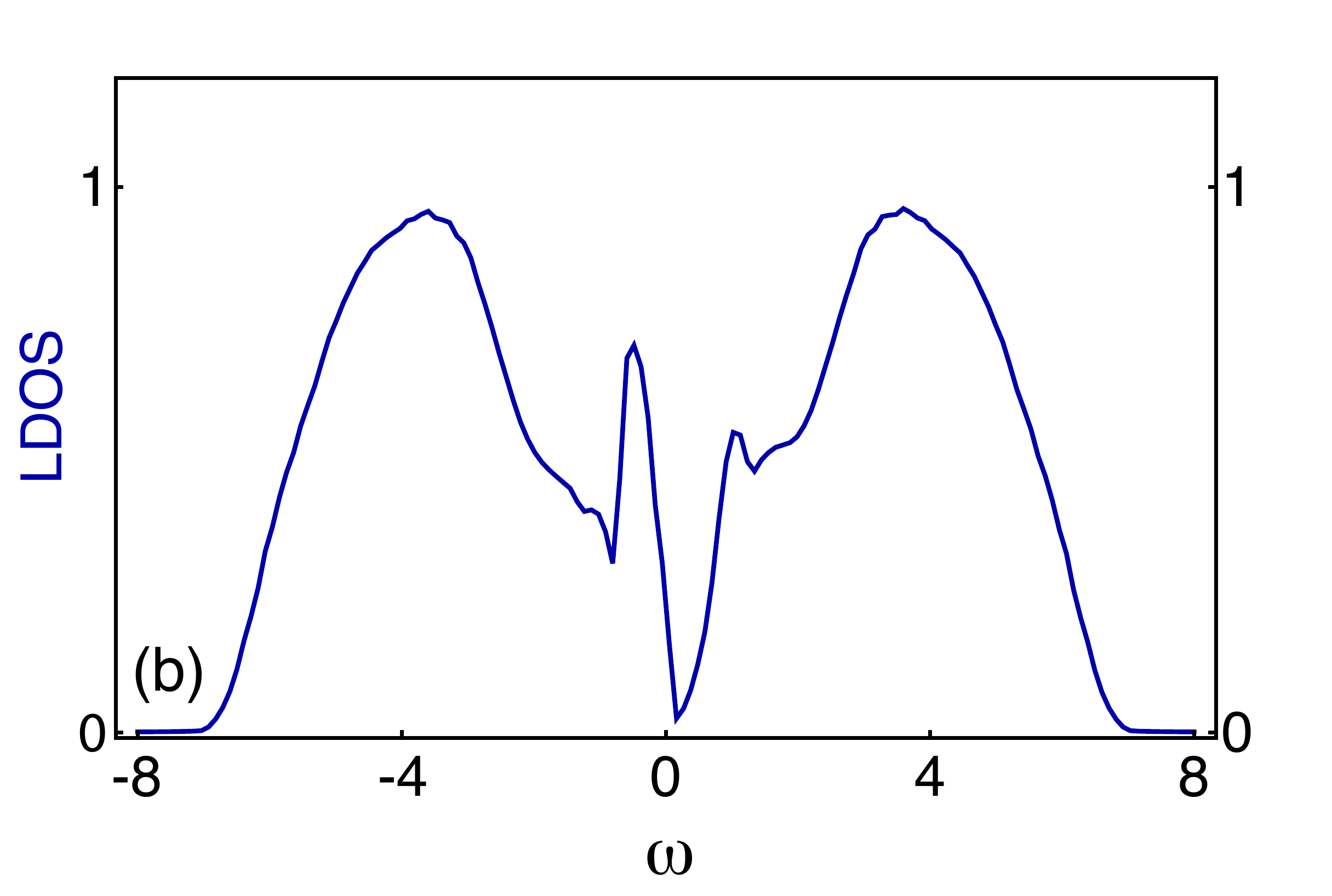}
      \end{center}
    \vspace{-0.5cm}
    \caption{
    (a) The contribution to the Berry phase
        for one spin component in a Dirac nodal line semimetal.
    (b) The local density of states  (LDOS) of the surface layer 
        Note: The above results are very similar in all cases for  
        a Dirac nodal line semimetal, a Weyl nodal line semimetal 
        with pure Rashba or pure Dresselhaus SOC, and  a Weyl nodal 
        line semimetal with mixed Rashba and Dresselhaus SOC. 
      }
       \label{Fig6}%
\end{figure}
%%%%%%%%%%%%%%%%%%%%%%%%%%%%%%%%%%%%%%%%%%%%%%%%%%%%%%%%%%%%%
%

The Berry phase picked up by a closed path is $\pi$ whenever the
nodal line is encircled and zero otherwise, showing the topological
charge of the nodal line.
We show the Berry phase calculated along $k_z$ in
Fig.~\ref{Fig6}(a).
More details on the calculation of the Berry phase are presented
in Appendix~\ref{APP_B}. 

%%%%%%%%%%%%%%%%%%%%%%%%%%%%%%%%Figure%%%%%%%%%%%%%%%%%%%%%%%%%%
\begin{figure}[t]
 \begin{center}
     \includegraphics[width=1 \linewidth]{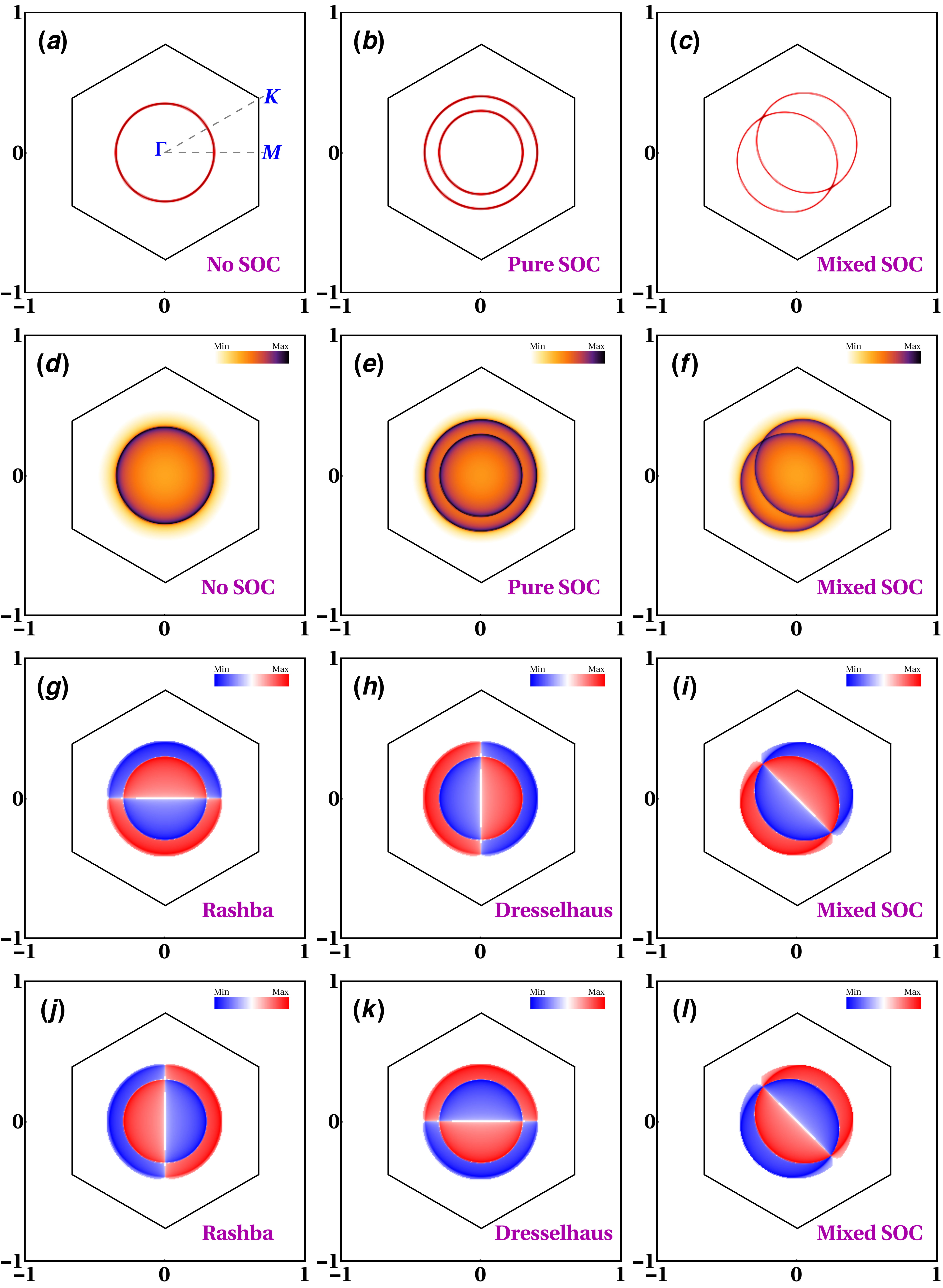}
\end{center}
\caption{
    Momentum  dependency of the spectral functions and their
    spin polarizations for 
    %the structure with  topological
    drumhead surface states.
    (a-c) Bulk Fermi surfaces within the $k_z=0$ correspond to the
    nodal rings.
    The subplots (d-f) show the momentum-resolved
    surface density of states
    $\rho_{n=1}(\bk_{\parallel},\omega=0)$.  The solid lines indicate the
    first BZ.
    In the first two rows, the left,
    middle and right rows represent (a,d) a DNLS, (b,e) a WNLS with pure
    Rashba/Dresselhaus, and (c,f) a WNLS with the same strengths of
    Rashba and Dresselhaus SOCs, respectively.
    The third and fourth rows indicate the $x$- and $y$- components of the
    spin-resolved surface spectral function
    $\rho^{i}_{n=1}(\bk_{\parallel},\omega=0)$ for the WNLS with (g,j)
    Rashba, (h,k) Dresselhaus, and  (i,l) mixed SOC. 
    The labels are $(k_x,~k_y)$ with units of $\pi/a$.  
  }
\label{Fig3}
\end{figure}
%%%%%%%%%%%%%%%%%%%%%%%%%%%%%%%%%%%%%%%%%%%%%%%%%%%%%%%%%%%%% 
%
%%%%%%%%%%%%%%%%%%%%%%%%%Figure%%%%%%%%%%%%%%%%%%%%%%%%%%
\begin{figure}[t]
 \begin{center}
          \hspace{-0.4cm}
              \includegraphics[width=0.95 \linewidth]{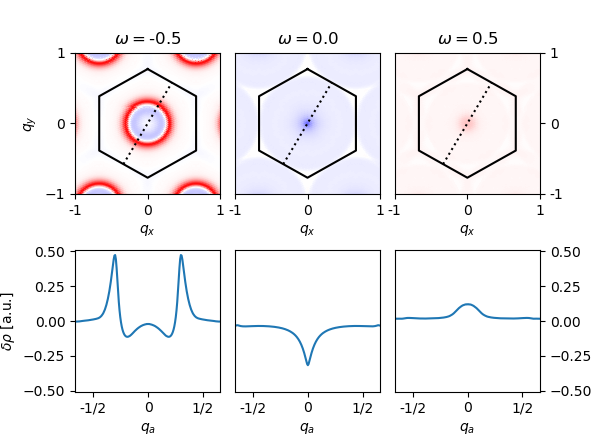}
    \vspace{-0.3cm}
\end{center}
\caption{ \label{mFig5}
    QPI patterns for the spinless case at different bias 
    $\omega$= -0.5, 0.0 and +0.5. The axis in the color plot
    are $(q_x,~q_y)$ in untis of $\pi/a$.
    The lower plot shows a cut along the dashed line.
    %Dirac nodal line semimetal     (a-c), Weyl nodal line
    %semimetal with pure Rashba/   Dresselhaus     (d-f), and Weyl
    %nodal line semimetal with admixture of Rashba and
    %Dresselhaus SOCs  (g-i).    
    %The left, middle and right columns denote $\omega=0$, $\pm
    %0.5$, and $-1$ in the unit of $(t_{\perp})$, respectively. 
    %The left column denotes $\omega=-1 (t_{\perp})$ and the right
    %represents  $\omega=0$, and the axis labels are  $(q_x,~q_y)$
    %with units of $\pi/a$.  
    }
\label{Fig4}
\end{figure}
%%%%%%%%%%%%%%%%%%%%%%%%%%%%%%%%%%%%%%%%%%%%%%%%%%%%%%%%%%%%% 
%
Now we show that  the nontrivial topology of this nodal ring
manifests itself as a so called drumhead surface state.  For this
purpose, we consider the system as a bundle of parallel
two-dimensional $xy$-slabs ($n=1,2,...,N_z$) piled upon one
another in the $z$-direction. The boundary condition is periodic
on each slab and open along the $z$-direction.  Using the partial
Fourier transformation
$c^{\dag}_{\bk,\gamma}=\frac{1}{N_z}\sum_{n} e^{{\mi} k_z z_{n}}
c^{\dag}_{\bk_{\parallel},n,\gamma}$, with
$\bk_{\parallel}=(k_x,k_y)$, the bulk BZ will be mapped into the
slab geometry configuration on the new basis
$
\Phi^{\dag}_{n,\bk_{\parallel}}
    =
(c^{\dag}_{n,\bk_{\parallel},d}, c^{\dag}_{n,\bk_{\parallel},p} )
$,
by the boundary condition
$\Phi_{0,\bk_{\parallel}}=\Phi_{N_z+1,\bk_{\parallel}}=0$ within
the two-orbital model, and $n=1,\dots,N_z$.  In this  basis the
Hamiltonian is given by~\cite{Lambert_2017,Zare_2017}
%
%===========================================================
%===================== Formula =============================
%===========================================================
\bea
\mathcal{H}_{\bk_\parallel}
    =
        \sum_{n=0}^{N_z+1} 
            \Phi^\dagger_{n,\bk_\parallel} \big( 
                 T          \Phi_{n-1,\bk_\parallel}
                +M          \Phi_{n  ,\bk_\parallel} 
                +T^\dagger  \Phi_{n+1,\bk_\parallel} 
            \big).
\label{H_slab}
\no\\
\eea
%
%===========================================================
%
The hopping matrix $T$ connects two consecutive slabs, while $M$
contains hopping terms within a slab as well as on-site energies.
Both the intra- and inter-orbital contributions are present  in
the hopping matrices $T$ and $M$, which are given by:
%
%===========================================================
%===================== Formula =============================
%===========================================================
\begin{align}
\begin{aligned}
M
=
&\:
{ \tilde{\varepsilon}_1(\bk_{\parallel})\htau_z - t^{\prime} \htau_x }
;\;\;
\\
T
=
&
-t_{\perp} \htau_z
-\frac{1}{2} t^{\prime} (\htau_x- {\mi} \htau_y).
\end{aligned}
\end{align}
%===========================================================
Here $\tilde{\varepsilon}_1$ is given by Eq.~\eqref{epsi_1}, but
without the $k_z$ dependent terms. 

For $\bk_{\parallel}$ within the projected nodal ring, there is a
band inversion and there are surface bound states among the
eigenstates of the slab Hamiltonian~\eqref{H_slab}, which decay
exponentially along the $z$-direction into the bulk.  These
surface states are a manifestation of the topological properties
of the system. 
To reveal these surface states it is useful to compute the local
density of states (LDOS), $N_n(\omega)$, for a given slab $n$.
The LDOS is computed by use of the Matsubara and retarded Green's
functions, which are defined for the slab geometry as
$\hat{G}(\bk_{\parallel},{\mi} \omega)=({\mi}
\omega-\cal{H}_{\bk_{\parallel}})\inv$ 
and
$\hat{G}^{ret}(\bk_{\parallel},\omega)=\hat{G}(\bk_{\parallel},
{\mi} \omega \rightarrow \omega + {\mi} 0^{+})$,
respectively. With this, the LDOS for the $n{\rm th}$ slab is
obtained as
%
%===========================================================
%===================== Formula =============================
%===========================================================
\be
N_n(\omega)=-\frac{1}{\pi}\sum_{\bk_{\parallel}} {\rm Im} 
\Big[ {\rm Tr} 
\Big [
 \hat{G}^{ret}_{nn}(\bk_{\parallel},\omega) 
 \Big ]
  \Big],
\label{DNSL_LDOS}
\ee
and the momentum-resolved LDOS reads
%===========================================================
%===================== Formula =============================
%===========================================================
\be
\rho_n(\bk_{\parallel},\omega)=-2 {\rm Im} 
\Big[ {\rm Tr} 
\Big[
 \hat{G}^{ret}_{nn}(\bk_{\parallel},\omega) 
\Big] 
\Big],
\label{DNSL_Spec}
\ee
%===========================================================
where the trace runs over orbital and spin degrees of freedom.
%===========================================================
 In Fig. \ref{Fig2}(d) we present the momentum-resolved spectral function
  (momentum-resolved density of states)
   for the surface slab of the DNLS along high-symmetry lines in the BZ. This reveals the formation of 
 topological surface states around the $\Gamma$ point, in the interior of the bulk Dirac nodal ring. 
Indeed, the surface modes can be easily observed as an in-gap drumhead surface state in the momentum-resolved spectral function of the surface slab inside the hexagonal BZ [see Fig. \ref{Fig3}(d)].  
 These two-dimensional drumhead surface states are bounded by the projection of the one-dimensional bulk Dirac nodal ring onto the surface.
Fig.~\ref{Fig6}(b) shows the $k$-integrated LDOS of the surface
layer.
While a nodal line semimetal has a vanishing DOS at Fermi level in
the bulk, the surface states contribute to a non-vanishing value.
The energy range of their dispersion is marked by Van Hove
singularities
% moved up from further down

%
%%%%%%%%%%%%%%%%%%%%%%%%%Figure%%%%%%%%%%%%%%%%%%%%%%%%%%
%\begin{figure*}[t]
 %\begin{center}
  %   \includegraphics[width=0.9 \linewidth]{Fig5.pdf}
%\end{center}
 %        \vspace{-0.5cm}
%\caption{
%Energy dispersion of the bulk  spin-resolved density of states along high-symmetry lines within the  $k_z=0$ plane of the BZ, for
 %  $x$-components (upper panels), and  $y$-components (lower panels). (a,d) the Weyl nodal line semimetal (WNLS) with pure Rashba SOC,   (b,e) the WNLS with pure Dresselhaus SOC, and (c,f) the WNLS with mixed Rashba and Dresselhaus SOC. 
  %  Along the $\Gamma\rightarrow$M path, the band structures of the WNLSs do not have $x$($y$)-components for Rashba (Dresselhaus) SOC, which can be easily understood from the Eq.~\eqref{Spin_Expectation}. 
   %
   %   }
%
    %   \label{Fig5}%
%\end{figure*}
%%%%%%%%%%%%%%%%%%%%%%%%%%%%%%%%%%%%%%%%%%%%%%%%%%%%%%%%%%%%%
%

Quasiparticle interference (QPI) is an efficient tool to observe
topological surface states in different classes of topological
materials~\cite{hofmann_QPI_topSC,lambert_eremin_QPI_Weyl}. It can
provide useful information about the topological properties of the
surface electronic band structure. QPI patterns can be
experimentally obtained from the Fourier transformation of
spatially modulated STM data, due to the elastic scattering of
quasiparticles by an impurity potential  $V$.
In the framework of the full Born approximation,
QPI 
for a point like impurity
is proportional to the variation of the LDOS and obtained
by~\cite{Akbari_QPI_2013,Akbari_t_matrix_2013}
%
%===========================================================
%===================== Formula =============================
%===========================================================
\be
{\rm QPI}=\delta N(\bq_{\parallel},\omega)=-\frac{V}{\pi}{\rm Im}[\Lambda(\bq_{\parallel},{\mi}\omega)]_{{\mi}\omega \rightarrow \omega+{\mi} 0^{+}}.
\label{QPI_LDOS_Modulation}
\ee
%===========================================================
%
Using the $T$-matrix formalism for a slab geometry
\cite{Lambert_2017}, the QPI function
$\Lambda(\bq_{\parallel},{\mi}\omega)$ is given by
%===========================================================
%===================== Formula =============================
%===========================================================
\be
\Lambda(\bq_{\parallel},{\mi}\omega)
    =
        \frac{1}{N}\sum_{\bk_{\parallel}} {\rm Tr} [ 
            \hat{\varrho}_0  \hat{G}(\bk_{\parallel}, {\mi}\omega) 
                        \hat{G} (\bk_{\parallel}-\bq_{\parallel},{\mi}\omega)].
%\no\\
\label{DNSL_QPI}
\ee
%===========================================================
Here, the interaction matrix 
$\hat{\varrho}_i = \tau_0\sigma_i$ acts on the orbital and spin
space and distinguishes charge and spin channels, $N$ is the
number of grid points. 
% Figure~\ref{Fig4}(g) displays   the intensity plot of the QPI
% pattern for the drumhead surface state of a DNLS.
%This figure should be compared to the corresponding
%momentum-resolved spectral function in Fig. \ref{Fig4}(d).
% We observe that the QPI pattern exhibits a logarithmic
% divergence at $| {\bf q}_{\parallel} | = 2 k_0$, i.e., along a
% ring with radius $2 k_0$, where $k_0$ is the radius of the bulk
% nodal ring.
%
% In particular,
%Figs.~\ref{Fig4}(a)-~\ref{Fig4}(c) display the intensity plots of
%the QPI patterns for the drumhead surface state of a DNLS with
%$\omega=0$, $\pm t_{\perp}/2,$ and $\pm t_{\perp}$, respectively.
 %
In Fig.~\ref{Fig4} we present the QPI at different energies. 
From (\ref{DNSL_QPI}) one can see that major contributions to the
sum are given when both Greens functions have poles simultaniously 
at $\bk_\parallel$ and $\bk_\parallel-\bq_\parallel$, i.e. 
$E_1(\bk_\parallel) \approx E_2(\bk_\parallel - \bq_\parallel) 
\approx \omega$.
Fig.~\ref{Fig4} should therefore be compared to the corresponding
momentum-resolved spectral function in Fig.~\ref{Fig3}(d) at the
given $\omega$.
As mentioned above, the non-vanishing LDOS at Fermi level is
due to the surface states, while the bulk contribution vanishes.
As a consequence, the QPI is dominated by the surface states
as well at those energies.

We observe that the QPI patterns  exhibit  
divergences
when $\bq_\parallel$ connects the surface states at the given
$\omega$ on opposite sites of the ring, as shown in the case 
$\omega=-0.5$.
The sharply defined dispersion of the surface states also results
in a sign change at the divergence, while the bulk state
contribution is smooth, since they form a continuum, e.g. at
$\omega=0.5$ where no surface states are present.
Note that the radius of the QPI pattern is twice the radius
of the ring at a given energy. As we approach $\omega = 0$ from
below, the divergences appears at larger $|\bq_\parallel|$
up to twice the radius of the nodal line.
At the same time, the surface states become less localized and 
merge with the bulk states eventually, such
that no clear signature from the surface states can be seen at
$\omega=0$. 
%For $\omega$ within the energy range of the dispersing surface
%states, the divergence is at twice the radius of the surface 
%states at that energy.

%In Figs.~\ref{Fig4}(c)-\ref{Fig4}(f) we show the intensity of the
%QPI patterns, $\delta N ( {\bf q}_{\parallel}, \omega)$,  for the
%drumhead surface states of WNLSs.
%Figs.~\ref{Fig4}(c)-\ref{Fig4}(d) corresponds to the QPI patterns
%for a WNLS with pure Rashba/Dresselhaus SOC, 
%where the spin split leads to two independent contributions
%at different radii, leading to the concentric rings.
%which clearly exhibit the $C_3$ symmetry of the crystal. 
%Interplay between Rashba and Dresselhaus SOCs results in a
%reduction in the symmetry,  which is clearly visible in the QPI
%patterns of Fig.~\ref{Fig4}(e)-\ref{Fig4}(f).  
%

While the surface state dispersion and therefore the precise
form of the QPI patterns depend on the band dispersion of the
material, the overall features should be stable under deformations.
Especially the sharp divergence and the sign change
is a telltale signature of the drumhead surface state, which can
be used in STM-based QPI experiments as a fingerprint of the
topological surface state.

%%%%%%%%%%%%%%%%%%%%%%%%%Figure%%%%%%%%%%%%%%%%%%%%%%%%%%
\begin{figure}[t]
 \begin{center}
     \includegraphics[width= 1\linewidth]{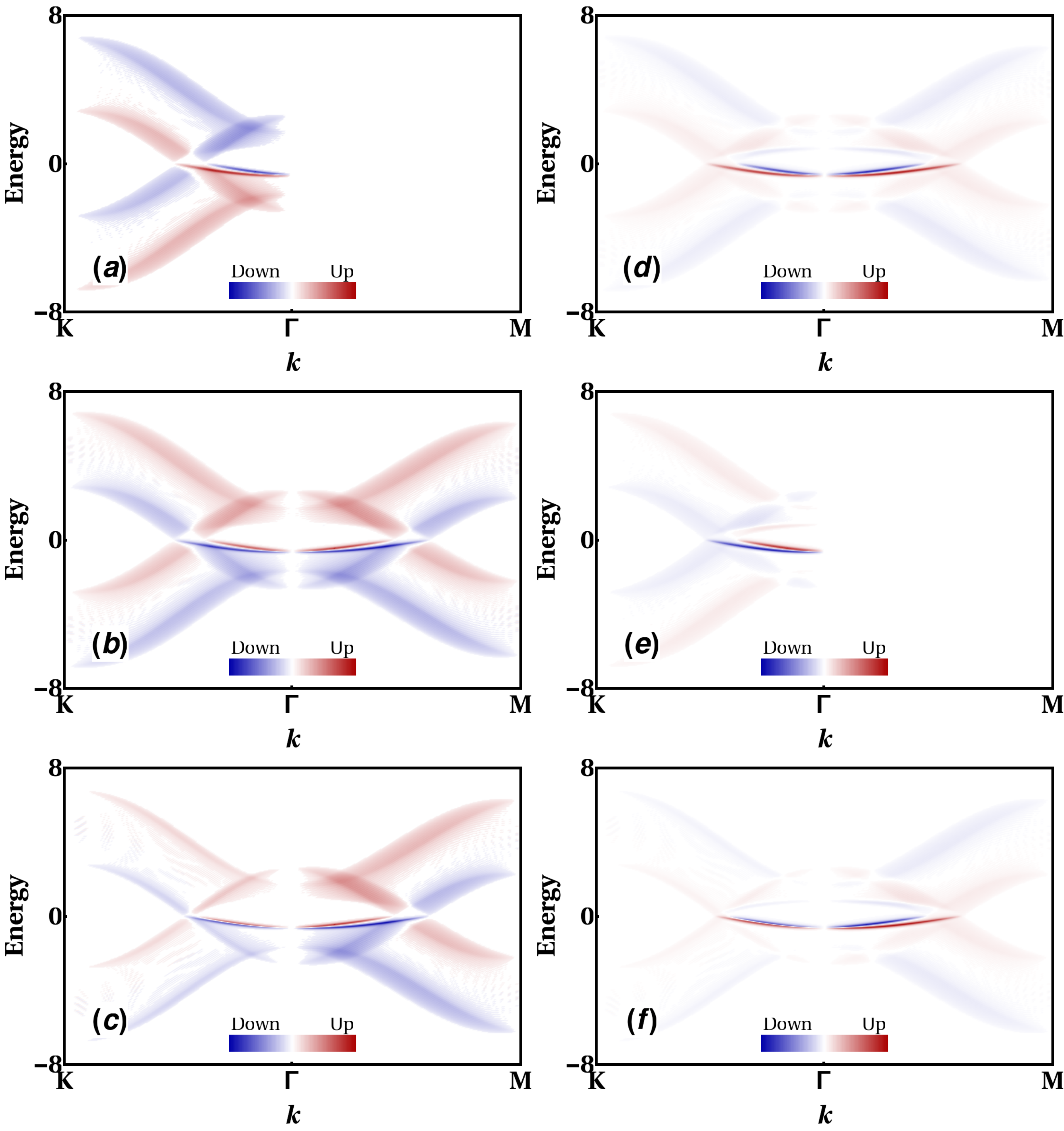}
\end{center}
    \vspace{-0.5cm}
\caption{
Energy dispersion of 
left: $x$-, and right: $y$-components of the spin-resolved ``{\it surface density of states}" along high-symmetry lines of the BZ for (a,d) the Weyl nodal line semimetal  (WNLS) with  pure Rashba SOC, 
 (b,e) the WNLS with pure Dresselhaus SOC, and (c,f) the WNLS with mixed Rashba and Dresselhaus SOC.
      }
       \label{Fig5}%
\end{figure}
%%%%%%%%%%%%%%%%%%%%%%%%%%%%%%%%%%%%%%%%%%%%%%%%%%%%%%%%%%%%%%%%%%
%%%%%
% 

%%%%%%%%%%%%%%%%%%%%%%%%%%%%%%%%%%%%%%%%%%%%%%%%%%%%%%%%%%%%% %
\subsection{Weyl nodal line semimetal with  antisymmetric SOC}
Let us now discuss how the lack of inversion symmetry affects the
topology of a nodal line semimetal.
Due to breaking of spatial
inversion symmetry through the additional atoms in the unit cell,
 space group of the candidate materials considered here is 
D$_{3h}$ (189).
%in noncentrosymmetric crystals, like TlTaSe$_2$
%and PbTaSe$2$,  
This allows for the spin degeneracy to be lifted,
resulting in a horizontal shift in $\bk$ space in opposite 
directions for bands of opposite spin upon inclusion of SOC.
%As a consequence of SOC, the bands with
%opposite spins shift horizontally in opposite directions in
%$\bk$-space. 
As a consequence, nodal-line systems without
inversion symmetry exhibit Weyl nodal rings with only two-fold
degeneracy, which are protected by reflection symmetry along the
$z$-direction.  Lack of inversion symmetry induces an
antisymmetric Rashba or Dresselhaus SOC.  The real space
Hamiltonian of the WNLS with Rashba and Dresselhaus SOCs is given
by
%===========================================================
%===================== Formula =============================
%===========================================================
\begin{align}
 \label{H_r_WNLS}
\begin{aligned}
{\cal H}
=
&
{\cal H}_0 
 +
\frac{{\mi} \: \alpha}{2} 
\sum_{\langle ij \rangle,\gamma\gamma',\sigma\sigma'}  
\!\!\!
\htau_z^{\gamma\gamma'}(\hat{\boldsymbol{\sigma}} \times \hat{r}_{ij})_{\sigma\sigma'} c^{\dag}_{i,\gamma,\sigma}c^{}_{j,\gamma',\sigma'}
\\
& +
\frac{{\mi} \: \beta}{2} \sum_{\langle ij \rangle,\gamma\gamma',\sigma\sigma'}  
\htau_z^{\gamma\gamma'}(\hat{\tilde{\boldsymbol{\sigma}}} \cdot \hat{r}_{ij})_{\sigma\sigma'} c^{\dag}_{i,\gamma,\sigma}c^{}_{j,\gamma',\sigma'} , 
\end{aligned}
\end{align}
%===========================================================
where $\alpha$ and $\beta$ are the strengths of Rashba and Dresselhaus SOC. Furthermore, $\sigma$ and $\sigma'$ are spin indices and 
$\hat{\tilde{\boldsymbol{\sigma}}}=(\hat{\sigma}_x,-\hat{\sigma}_y)$. 
In momentum space,
the Hamiltonian  
 is given by
%
%===========================================================
%===================== Formula =============================
%===========================================================
\begin{align}
\begin{aligned}
{\cal H}=&
{\cal H}_0 
 +
 \alpha
 \!\!
 \sum_{\bk,\gamma\gamma',\sigma\sigma'}~ \htau_z^{\gamma\gamma'}(\bg^R_{\bk} \cdot 
 \hat{\boldsymbol{\sigma}})_{\sigma\sigma'} c^{\dag}_{\bk,\gamma,\sigma}c^{}_{\bk,\gamma',\sigma'}
\\
& +
 \beta
  \!\!
  \sum_{\bk,\gamma\gamma',\sigma\sigma'} ~ \htau_z^{\gamma\gamma'}(\bg^D_{\bk} \cdot 
 \hat{\boldsymbol{\sigma}})_{\sigma\sigma'} c^{\dag}_{\bk,\gamma,\sigma}c^{}_{\bk,\gamma',\sigma'}  ,
 \label{k_space_Ham}
\end{aligned}
\end{align}
%===========================================================
where $\bg^R_{\bk}$ and $\bg^D_{\bk}$ are Rashba and Dresselhaus spin-orbit $\bg$-vectors and are defined as
%===========================================================
%===================== Formula =============================
%===========================================================
\begin{align*}
\begin{aligned}
\bg^R_{\bk}=
&
\sqrt{3}a  \sin(\frac{\sqrt{3}k_ya}{2}) \Big[\cos(\frac{3k_xa}{2})+2\cos(\frac{\sqrt{3}k_ya}{2}) \Big]~\hat{\bf x} 
\\
&
- 3a \sin(\frac{3k_xa}{2})\cos(\frac{\sqrt{3}k_ya}{2})~\hat{\bf y},
\end{aligned}
\end{align*}
and
\begin{align*}
\begin{aligned}
\bg^D_{\bk}=
&
3a \sin(\frac{3k_xa}{2})\cos(\frac{\sqrt{3}k_ya}{2})~\hat{\bf x}  
\\
&
- \sqrt{3}a \sin(\frac{\sqrt{3}k_ya}{2}) \Big[\cos(\frac{3k_xa}{2})+2\cos(\frac{\sqrt{3}k_ya}{2})\Big]~\hat{\bf y}.
\end{aligned}
\end{align*}
%===========================================================
Under the operation of parity ($\bk \rightarrow -\bk$) both spin-orbit ${\bf g}$-vectors  are antisymmetric, i.e., $\bg^{R/D}_{-\bk}=-\bg^{R/D}_{\bk}$.
Moreover, Eq.~(\ref{k_space_Ham}) shows that Rashba and Dresselhaus 
SOCs play the role of momentum-dependent Zeeman fields,  thereby splitting the
spin degeneracy. Since SOC in this form mixes only states of equal 
$R_z$ eigenvalues, the line node is still protected by reflection symmetry.

The energy dispersions of the bulk Hamiltonian~\eqref{k_space_Ham} at $k_z=0$ along the K$\Gamma$MK-path in the presence of a pure Rashba SOC, pure  Dresselhaus SOC, and 
an equal combination of both Rashba and Dresselhaus SOCs are shown in 
Figs.~\ref{Fig2}(b) and  Figs.~\ref{Fig2}(c). 
Due to the presence of antisymmetric SOC, the bulk bands are now spin split. 
This splitting can be observed in Fig.~\ref{Fig3}(b),
where two separate and concentric Weyl nodal rings are present. 
Our calculations yield identical results for the case of pure Rashba SOC and pure Dresselhaus SOC. However, mixing of Rashba and Dresselhaus SOC leads to an anisotropic nodal structure in the BZ. 
 We show this in Fig.~\ref{Fig3}(c) for $\alpha=\beta$ (i.e., equal strengths of Rashba and Dresselhaus SOCs), in which case the two nodal rings intersect each other along the $[1\bar{1}]$ direction. 
This anisotropy can also be seen in the asymmetry of the dispersion 
in K$ \rightarrow \Gamma$ and $\Gamma \rightarrow$ M directions 
in Fig.~\ref{Fig2}(c).
\\

With the same approach as for the case of DNLSs, we investigate 
the surface states and the topological properties of WNLSs  using 
an $xy$-slab geometry. However, for WNLSs, we now need to take 
the spin degrees of freedom into account .
The slab Hamiltonian for the WNLSs is of the same form as Ref.~[\ref{H_slab}],
but with the following spin-dependent hopping matrices
%
%===========================================================
%===================== Formula =============================
%===========================================================
\begin{align}
\begin{aligned}
M
\!
=&
\;
{\left[ \tilde{\varepsilon}_1(\bk_{\parallel})\htau_z 
        -t^{\prime} \htau_x 
    \right] \hat{\sigma}_0 }
+[   \alpha{\rm g}^R_x(\bk_{\parallel})
    +\beta {\rm g}^D_x(\bk_{\parallel}) ] \htau_z \hat{\sigma}_x
\\ 
&+[  \alpha{\rm g}^R_y(\bk_{\parallel})
    +\beta {\rm g}^D_y(\bk_{\parallel}) ] \htau_z \hat{\sigma}_y,
\\
\vspace{0.1cm}
\\
T
=
&
-\left[ 
    t_{\perp} \htau_z 
    +\frac{1}{2} t^{\prime} (\htau_x- {\mi} \htau_y)
    \right]\hat{\sigma}_0,
\end{aligned}
\end{align}
%============================================================
 where $\hat{\sigma}_i$ operates in spin space
and ${\rm g}_x^{R/D}$ and ${\rm g}_y^{R/D}$ are $x$- and $y$- components of the Rashba/Dresselhaus $\bg$-vectors. 
 Figs.~\ref{Fig2}(e) and~\ref{Fig2}(f) portray the momentum-resolved spectral function for the surface slab of the WNLS with pure Rashba/Dresselhaus 
 and their combination with equal strengths. The horizontal shifts in the band structures due to SOC is clearly observed. In addition, 
 one can see that in the presence of both Rashba and Dresselhaus SOC, the topological surface states become anisotropic. 
In Figs.~\ref{Fig3}(e) and ~\ref{Fig3}(f) we present the
momentum-resolved spectral functions at zero energy, $\omega=0$,
for  the surface slab  of the WNLSs. Figure~\ref{Fig3}(e)
corresponds to a WNLS with pure Rashba/Dresselhaus SOC. Here, two
isotropic and concentric drumhead surface states are observed.  In
Fig.~\ref{Fig3}(f), one can see two shifted drumhead surface
states of equal width, which stem from an interplay between Rashba
and Dresselhaus SOC with equal strengths.

%%%%%%%%%%%%%%%%%%%%%%%%%%%%%%%%%%%%%%%%%%%%%%%%%%%%%%%%%%%%%%%%%%
\subsection{Spin texture of Weyl nodal line semimetals}

Antisymmetric SOC splits the spin degeneracy of the bands, thereby
creating a nontrivial spin texture of the bulk and surface band
structure of WNLSs.  Therefore, compared to a DNLS, the band
structure in a WNLS is strongly spin polarized. Due to the locking
of orbital and spin degrees of freedom by SOC, none of the spin
and momentum can be supposed as well-defined quantum numbers.
Consequently the Hamiltonian doesn't remain diagonal in spin
space.  For diagonalizing the Hamiltonian of a WNLS, one has to
change the basis to the helicity basis (${|\pm \rangle}$). In the
helicity basis, the expectation values of the different components
of the spin operator, $\hat{S}^{i}$, are given by
%=================================================================
%===================== Formula ===================================
%=================================================================
\be
\langle \hat{S}^{i} \rangle_{\pm}=
\langle {\pm}| \hat{S}^{i} |{\pm}\rangle
=
 {\pm}\hat{\bg}^{i}_{\bk},
\label{Spin_Expectation}
\ee
%=================================================================
with $\hat{\bg}^{i}_{\bk}=\bg^{i}_{\bk}/|\bg_{\bk}|$. 
%
%Figures~\ref{Fig5}(a)-\ref{Fig5}(f) show the $x$- and $y$-components of the spin-resolved band structure along high-symmetry lines within the $k_z=0$ plane of the BZ for (a,d) the WNLS   with pure Rashba SOC,  (b,e) the WNLS with pure Dresselhaus SOC, and (c,f) the WNLS with mixed   Rashba and Dresselhaus SOC.
%
Along the $\Gamma\rightarrow{\rm M}$~path the momentum component $k_y$ is zero, $k_y=0$, and consequently
Eq.~(\ref{Spin_Expectation}) implies that
 $\hat{\bg}^{x}_{\bk}$ and $\hat{\bg}^{y}_{\bk}$ are zero in a WNLS with pure Rashba and pure Dresselhaus SOC, respectively.
Therefore
the $x$-component ($y$-component) of the spin polarization of  the
WNLS with pure Rashba (pure Dresselhaus) SOC is zero in the
above-mentioned direction.  

In the same way as in the bulk, the spin degeneracy of the 
surface states of the WNLS is lifted and they become spin 
polarized.
To examine the spin 
texture of the topological surface states, we are using the 
spin-resolved spectral function. 
For the $n{\rm th}$ slab it is defined as
%===========================================================
%===================== Formula =============================
%===========================================================
\be
\rho^{i}_n(\bk_{\parallel},\omega)=-2 {\rm Im} \Big[ {\rm Tr} \Big[ \hat{S}^{i} \hat{G}^{ret}_{nn}(\bk_{\parallel},\omega) \Big] \Big],
\label{DNSL_Spec}
\ee
%===========================================================
where $\hat{S}^{i}=\tau_0\frac{1}{2}\sigma_i$.
As seen in Figs.~\ref{Fig3}(g)-\ref{Fig3}(i), antisymmetric Rashba and
Dresselhaus SOC generate two concentric spin-polarized drumhead
surface states, whose total spin polarizations are inside the
$xy$-plane. One can clearly observe that the Weyl nodal rings are fully
spin-polarized in opposite directions.
Figures~\ref{Fig5}(a-f) show the $x$- and $y$-components of the
spin-resolved spectral functions along the K$\Gamma$MK-path for
the first slab of a WNLS with (a,d) pure Rashba SOC, (b,e) pure
Dresselhaus SOC,  and (c,f) mixed Rashba and Dresselhaus SOC.
% One can clearly observe that the Weyl nodal rings are fully
% spin-polarized in opposite directions. As seen in
% Figs.~\ref{Fig4}(j)-\ref{Fig4}(o), antisymmetric Rashba and
% Dresselhaus SOC generate two concentric spin-polarized drumhead
% surface states, whose total spin polarizations are inside the
% $xy$-plane.
%

In order to investigate the effect of SOC via QPI, we have to
move from non-magnetic to magnetic impurities.
Scattering at non-magnetic impurities allows for scattering
between states of the same spin orientation. From the spin
polarization of the surface states it becomes clear, that the
contribution from the surface states is given by inter-surface
state scattering only. Therefore, the shift in the radius of 
the two nodal lines is not visible and the pattern looks almost
identical to the spinless case.

\begin{figure}
\includegraphics[width=.9\linewidth]{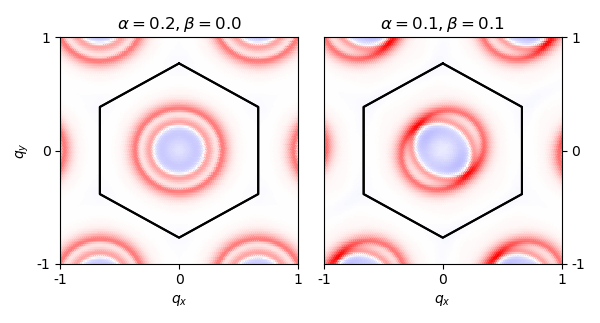}
\caption{ \label{mFig7}
    Spin resolved QPI at $\omega=-0.5$ for a magnetic impurity
    with the magnetic moment pointing along $z$ for 
    (a) Rashba or Dresselhaus SOC and 
    (b) the case of combined Rashba and Dresselhaus SOC.
    The axes are spanned by $(q_x,q_y)$ in units of $\pi/a$.
    }
\label{QPI_spinful}
\end{figure}

For a magnetic impurity in $z$-direction, which can be controlled
by applying a small magnetic field, the scattering involves a spin
flip for all states, since they are all in the $xy$-plane.
For this case, Eq.~\eqref{DNSL_QPI} in the calculation of the 
QPI pattern contains the spin matrix of the impurity and is
modified to
%===========================================================
%===================== Formula =============================
%===========================================================
\be
\Lambda^{(3,3)}(\bq_{\parallel},{\mi}\omega)
    =
        \frac{1}{N}\sum_{\bk_{\parallel}} {\rm Tr} [ 
            \hat{\varrho}_3 \hat{G}(\bk_{\parallel}, {\mi}\omega) 
                            \tau_0\sigma_3
                            \hat{G} (\bk_{\parallel}-\bq_{\parallel},{\mi}\omega)].
%\no\\
\label{SPIN_POLARIZED_QPI}
\ee
The spin resolved QPI in $z$-direction for such an impurity is
sensitive to the intra-band scattering only, while all inter-band
contributions are suppressed now.   
We present the patterns for the above mentioned cases in 
Fig.~\ref{QPI_spinful}. Each of the surface states contributes
their own divergence when $\bq$ equals the diameter, leading to
the double ring structure.

%%%%%%%%%%%%%%%%%%%%%%%%%%%%%%%%%%%%%%%%%%%%%%%%%%%%%%%%%%%%%%%%%%
%%%%%%%%%%%%%%%%%%%%%%%%%%%%%%%%%%%%%%%%%%%%%%%%%%%%%%%%%%%%%%%%%%
\section{CONCLUSIONS}
In this paper, we have studied the topological surface states of
Dirac and Weyl nodal-line semimetals, whose nodal lines are
protected by reflection and time-reversal symmetry.  In the
presence of spatial inversion symmetry, our two-orbital
tight-binding model corresponds to a Dirac nodal line semimetal
with a four-fold degenerate band crossing, supporting the
realization of  spin degenerate drumhead surface states.  In the
absence of inversion symmetry, by including antisymmetric Rashba
and Dresselhaus SOC, our model  describes a topological Weyl
semimetal with  two-fold degenerate nodal rings. We have shown
that  pure Rashba and pure Dresselhaus SOC have the same effects
on the topology of the electronic band structure and the surface
states.  In addition, we have found that in the presence of mixed
Rashba and Dresselhaus SOC, the Weyl nodal rings become
anisotropic.  Since antisymmetric SOC lifts the spin degeneracy,
the drumhead surface states of Weyl nodal line semimetals are
strongly spin-polarized with a spin polarization vector within the
$xy$-plane. By evaluating the Berry phase, we have demonstrated
 that the drumhead surface states of the Dirac and Weyl nodal-line semimetals arise from the non-trivial topology of the band
crossing in the bulk.

 We have also computed the quasiparticle interference (QPI) patterns for
the surface of Dirac and Weyl nodal-line semimetals and have
shown that these interference patterns contain unique
signatures of the drumhead surface states. 
Namely,  
for the Dirac nodal-line semimetal there appears a single ring,
both in the ordinary and in the spin-resolved QPI patterns (Fig.~\ref{mFig5}). However, 
for the Weyl nodal-line semimetal there appear two rings in the spin-resolved QPI pattern, see Fig.~\ref{mFig7}.
We have checked that these features do not depend on the particular parameter choice.
They are therefore generic to any nodal-line material and should be observable  in Fourier-transform scanning tunnelling spectroscopy.
The wavelength of the Friedel oscillations,  giving rise to the ring in the QPI pattern, depends
on the size of the Dirac or Weyl nodal ring in the bulk. 
 For example, in CaAgAs the nodal ring has a radius of about 0.1 \AA$^{-1}$~\cite{yamakage_2016}, while in PbTaSe$_2$ it is about 0.2 \AA$^{-1}$~\cite{bian_2016}. 
 Hence, the Friedel oscillations have a wavelength of about 10 -- 30 \AA, which should be measurable in STM experiments. 
With regards to energy resolution, we find that the gap within which the drumhead state exists, is typically of the order of a few 100 meV 
and the SOC is typically about 50-100 meV~\cite{bian_2016,bian_a_2016,yamakage_2016}. These energy scales are easily resolvable with STM.
We hope that our findings will stimulate such experiments.

%%%%%%%%%%%%%%%%%%%%%%%%%%%%%%%%%%%%%%%%%%%%%%%%%%%%%%%%%%%%%%%%%%
%%%%%%%%%%%%%%%%%%%%%%%%%%%%%%%%%%%%%%%%%%%%%%%%%%%%%%%%%%%%%%%%%%
\section*{Acknowledgments}
We thank A.~Bangura, and P. Thalmeier for  valuable comments and
illuminating discussions. M.\ B. is grateful to  I. De Marco, F.
Cossu,  H. Yavari and M. H. Zare for useful discussions.  This
work was supported  through  NRF funded by MSIP of Korea
(2015R1C1A1A01052411) and (2017R1D1A1B03033465), and by  Max
Planck POSTECH / KOREA Research Initiative (No. 2011-0031558)
programs through NRF funded by MSIP of Korea. 
M.\ B. acknowledges the receipt of the grant No. AF-03/18-01 from Abdus Salam International Center for Theoretical Physics, Trieste, Italy.
%
%

%%%%%%%%%%%%%%%%%%%%%%%%%%%%%%%%%%%%%%%%%%%%%%%%%%%%%%%%%%%%%%%%%%
%%%%%%%%%%%%%%%%%%%%%%%%%%%%%       appendices       %%%%%%%%%%%%%
\appendix
%%%%%%%%%%%%%%%%%%%%%%%%%%%%%%%%%%%%%%%%%%%%%%%%%%%%%%%%%%%%%%%%%%

%%%%%%%%%%%%%%%%%%%%%%%%%%%%%%%%%%%%%%%%%%%%%%%%%%%%%%%%%%
\section{Theory of QPI for a semimetal}
Here we develop the theory of QPI for a Dirac or Weyl nodal line
semimetal in the presence of particle-hole symmetry.
We employ the Matsubara Green's function in a generalized orbital
and spin space, which is described by the field operator
$\psi^{\dagger}_{\bk} = ( 
    c^{\dagger}_{\bk,d,  \uparrow},
    c^{\dagger}_{\bk,d,\downarrow},
    c^{\dagger}_{\bk,p,  \uparrow},
    c^{\dagger}_{\bk,p,\downarrow} )$.
The Green's function on this space is given by
%
%=================================================================
%===================== Formula ===================================
%=================================================================
\begin{equation}
\hat{G}(\bk,{\mi} \omega)=
\begin{bmatrix}
\hat{\cal G}^{+}(\bk,{\mi} \omega) && \hat{\cal F}(\bk,{\mi} \omega)
\\
\hat{\cal F}^{\dagger}(\bk,{\mi} \omega) && \hat{\cal G}^{-}(\bk,{\mi} \omega)
\end{bmatrix},
\end{equation}
%=================================================================
where in  a  $2 \times 2$ spin space representation, 
$\hat{\cal G}^{\pm}(\bk,{\mi} \omega)$
and 
$\hat {\cal F}(\bk,{\mi} \omega)$ 
are Matsubara Green's functions for  intra- and inter-orbital hopping,
respectively.
 In the presence of spin-orbit coupling, the bare Green's functions  are expressed as
%=================================================================
%===================== Formula ===================================
%=================================================================
\begin{align}
\begin{aligned}
\hat{\cal G}^{\pm}(\bk,{\mi} \omega)
&=
\frac{1}{2} \sum_{\xi=\pm 1}
 (\hat{\sigma}_0+\xi \hat{\bg}_{\bk} \cdot \hat{\boldsymbol{\sigma}})
{\cal G}^{\pm}_{\xi}(\bk,{\mi} \omega),
  \\
 \hat{\cal F}(\bk,{\mi} \omega)
&=
\frac{1}{2} \sum_{\xi=\pm 1}
 {\cal F}_{\xi}(\bk,{\mi} \omega)
 \hat{\sigma}_0,
 \end{aligned}
\end{align}
%=================================================================
where $\hat{\rm \bf g}_{\bk}=\bg_{\bk}/ |\bg_{\bk}|$ with
%=================================================================
%======================Equation===================================
%=================================================================
\begin{align}
\begin{aligned}
 {\cal G}^{\pm}_{\xi}(\bk,{\mi} \omega)
 &=
 \frac{
 {\mi}\omega
 \pm
 \varepsilon_1(\bk)
 \pm \xi
 |{\rm \bg}_{\bk}|
 }
 {
  ({\mi}\omega)^2-E^2_{\bk\xi} 
 },
 \\
 {\cal F}_{\xi}(\bk,{\mi} \omega)
 &=
 \frac{
 -{\mi}\varepsilon_2(\bk)
 }
 {
  ({\mi}\omega)^2-E^2_{\bk\xi} 
 },
 \end{aligned}
 \end{align}
%=================================================================
%
 in which 
%=================================================================
%======================Equation===================================
%=================================================================
\begin{equation}
E_{\bk\xi}
=
\sqrt{
\Big(
\varepsilon_1(\bk)+\xi |{\rm \bg}|
\Big)^2
+
\varepsilon^2_2(\bk)
}
\end{equation}
%
 %==========================================================
  is the energy dispersion with helicity $\xi$. 
The modulation of LDOS %local density of state
 is calculated by taking the effect of scattering from magnetic or 
nonmagnetic impurities into account.
The impurity Hamiltonian has the form
%===========================================================
%===================== Formula =============================
%===========================================================
\begin{equation}
{\cal H}_{\rm imp}=\sum_{\bk\bq\delta} V(\bq)\psi^{\dag}_{\bk+\bq} \hat{\varrho}^{\delta} \psi_{\bk},
\end{equation}
%===========================================================  
where 
$\lbrace \hat{\varrho}^{\delta} \rbrace=\tau_0\sigma_\delta$
is the spin of the impurity site. 
The case $\delta=0$ corresponds to scattering from a nonmagnetic impurities and the cases $\delta (=x,y,z) $ are related to scattering via magnetic impurities. 
The changes in the STM tunneling conductance due to impurity scattering can be evaluated by
%===========================================================
%===================== Formula =============================
%===========================================================
\begin{align}
\begin{aligned}
\frac{d(\Delta I_{\delta})}{dV}
&
 \sim \Delta N_{\delta\delta'}(\br,\omega=V)
\\
&
=
-\frac{1}{\pi} {\rm Im}
\Big[
{\rm Tr}_{\sigma} \hat{\varrho}_{\delta} \Delta \hat{G}_{\delta'}(\br,\br,\omega)
\Big],
\end{aligned}
\end{align}
%===========================================================  
where, $\Delta \hat{G}_{\delta'}$ is the change of $4 \times 4$ 
matrix of Green's function due to impurity scattering from a
charges  $(\delta'=0)$ or magnetic $(\delta'=x,y,z)$ impurity.

It is common to define QPI as  the Fourier transformed  modulation 
of the LDOS from the scattering at an impurity,
\begin{widetext}
 %===========================================================
%===================== Formula =============================
%===========================================================
\begin{align}
\begin{aligned}
\Delta N_{\delta\delta'}(\bq,\omega)
    &=
        -\frac{1}{\pi} \sum_{\br} {\rm e}^{\mi \bq\cdot \br}
            {\rm Tr}_{\sigma}  \frac{1}{2\mi} \left[
                 \hat{\varrho}_{\delta} \Delta \hat{G}_{\delta'}(\br,\br,{\mi}\omega)
                 -\left(\hat{\varrho}_{\delta} \Delta \hat{G}_{\delta'}(\br,\br,{\mi}\omega)\right)^* \right]_{{\mi} \omega \rightarrow \omega+{\mi} 0^{+}} \\
       &=
        -\frac{1}{2\pi\mi} {\rm Tr}_{\sigma} \left[ \hat{\varrho}_\delta   \Delta \hat{G}_{\delta'}( \bq,{\mi}\omega) 
                                -\hat{\varrho}_\delta^* \Delta \hat{G}_{\delta'}^*(-\bq,{\mi}\omega) \right]_{{\mi} \omega \rightarrow \omega+{\mi} 0^{+}}.  \\
\end{aligned}
\end{align}
%=================================================================

In weak scattering limit, by ignoring the possibility of bound state formation, the differential conductance is stated~as
%=================================================================
%===================== Formula ===================================
%=================================================================
\begin{equation}
\Delta N_{\delta\delta'}(\bq,\omega)
=
-\frac{1}{2\pi{\mi}}V_{\delta'}(\bq) \Big[   \Lambda_{\delta\delta'}( \bq,{\mi} \omega)
                                                        -\Lambda_{\delta\delta'}^*(-\bq,{\mi} \omega)  
                                                \Big]_{{\mi} \omega \rightarrow \omega+{\mi} 0^{+}},
\label{appendixN}
\end{equation}
%=================================================================
%
%\end{widetext}
%
where
%=================================================================
%===================== Formula ===================================
%=================================================================
\begin{align}
\begin{aligned}%\no
\Lambda_{\delta\delta'}(\bq,{\mi} \omega)
&=
\frac{1}{N}\sum_{\bk} {\rm Tr}_{\sigma} \Big[ 
        \hat{\varrho}_{\delta} \hat{G}_0(\bk,{\mi} \omega) 
        \hat{\varrho}_{\delta'} 
        \hat{G}_0(\bk-\bq,{\mi} \omega) \Big],
\end{aligned}
\end{align}
%=================================================================
and N is the number of grid points and $\hat{G}_0$ is the Greens
function of the unperturbed system. 
When $\Lambda_{\delta\delta'}$ is symmetric in $\bq$, 
the bracket in Eq.~(\ref{appendixN}) reduces to 
${\rm Im}\Lambda_{\delta\delta'}(\bq,\mi\omega)$.
Finally, the QPI function $\Lambda_{\delta\delta'}$ is obtained as 
\cite{Akbari_QPI_2013,Akbari_t_matrix_2013}

%=================================================================
%===================== Formula ===================================
%=================================================================
\begin{align}
\begin{aligned}
\Lambda_{00}(\bq,{\mi} \omega)
=&
\frac{1}{4N}\sum_{\bk\xi\xi'}
\Big[
\Big(
1+\xi\xi'(\hat{\bg}_{\bk} \cdot \hat{\bg}_{\bk-\bq})
\Big)
 K^{\bk\bq}_{\xi\xi'}({\mi} \omega)
+
K'^{\bk\bq}_{\xi\xi'}({\mi} \omega)
\Big];
\\
\hspace{0.74cm}
%
%\\
\Lambda_{ii}(\bq,{\mi} \omega)
=
&
\frac{1}{4N}\sum_{\bk\xi\xi'}
\Big[
\Big(
1-\xi\xi'(\hat{\bg}_{\bk} \cdot \hat{\bg}_{\bk-\bq}-2 \hat{\rm g}^i_{\bk}\hat{\rm g}^i_{\bk-\bq}
\Big)
%\\
%&
 K^{\bk\bq}_{\xi\xi'}({\mi} \omega)
 +
 K'^{\bk\bq}_{\xi\xi'}({\mi} \omega)
 \Big],
\end{aligned}
\end{align}
%=================================================================
with the integration kernels
%=================================================================
%===================== Formula ===================================
%=================================================================
\begin{align}
\begin{aligned}
K^{\bk\bq}_{\xi\xi'}({\mi} \omega)
=&
\frac{
[ 
{\mi}\omega+\varepsilon_1(\bk)+\xi|\bg_{\bk}|
]
[
{\mi}\omega+\varepsilon_1(\bk-\bq)+\xi'|\bg_{\bk-\bq}|
]
     }
     {
     [({\mi} \omega)^2-
     E^2_{\bk\xi}]
     [ ({\mi} \omega)^2-E^2_{\bk-\bq\xi'}]
     };
%\\
\:\:\:\:\:\:
K'^{\bk\bq}_{\xi\xi'}({\mi} \omega)
=&
\frac{
\varepsilon_2(\bk)
\varepsilon_2(\bk-\bq)
     }
     {
     [({\mi} \omega)^2-
     E^2_{\bk\xi}]
     [ ({\mi} \omega)^2-E^2_{\bk-\bq\xi'}]
     }.     
\end{aligned}
\end{align}
%=================================================================
\end{widetext}

%%%%%%%%%%%%%%%%%%%%%%%%%%%%%%%%%%%%%%%%%%%%%%%%%%%%%%%%%%%%%%%%%%
%%%%%%%%%%%%%%%%%%%%%%%%%%%%%%%%%%%%%%%%%%%%%%%%%%%%%%%%%%%%%%%%%%
\section{Berry phase }
\label{APP_B}
The Berry phase is obtained through the eigenvalues of the Wilson
loop matrix for a closed path enclosing the nodal ring. For a
closed loop, the Berry phase ${\cal P}$ is given by
%=================================================================
%===================== Formula ===================================
%=================================================================
\be
{\cal P}(k_x,k_y)={\mi} 
\ln[\det(W_{\cal L}(k_x,k_y))],
\label{Berry_Phase}
\ee
%=================================================================
where the  discrete Wilson loop matrix is given by the product 
of the link matrices, $U_{n}$, made up by the occupied bands 
$|u_i (\bk_n)\rangle$ at each step $\bk_n \rightarrow \bk_{n+1}$ 
along the
closed path $\bk_0 \rightarrow \bk_N = \bk_0 + \bG$, with $\bG$
 as the lattice translation vector of the
reciprocal lattice,
%=================================================================
%===================== Formula ===================================
%=================================================================
\be
W_{\cal L}
=
\prod_{n=0}^{N-1} \frac{U_n}{|\det(U_n)|} ,
\label{Berry_Phase}
\ee
by the link matrix's elements defined as 
\be
U^{ij}_n
=
\langle u_i(\bk_n)|u_j(\bk_{n+1}) \rangle.
\ee
%=================================================================
%
Whenever the Berry phase ${\cal P}(k_x,k_y)\neq 0$, a nontrivial
in-gap state appears for $(k_x,k_y)$ at the surface BZ. 
The non-trivial Berry phase for band crossing around $\Gamma$ in
BZ guarantees the stability of topological drumhead surface states
[See Fig.~\ref{Fig6}(a)].

%%%%%%%%%%%%%%%%%%%%%%%%%%%%%%%%%%%%%%%%%%%%%%%%%%%%%%%%%%%%%%%%%%

%%%%%%%%%%%%%%%%%%%%%%%%%%%%%%%%%%%%%%%%%%%%%%%%%%%%%%%%%%%%%%%%%%
%%%%%%%%%%%%%%%%%%%%%%%%%%%% bibliography %%%%%%%%%%%%%%%%%%%%%%%%
\bibliographystyle{prsty}
\bibliography{ref_topology}
%%%%%%%%%%%%%%%%%%%%%%%%%%%%%%%%%%%%%%%%%%%%%%%%%%%%%%%%%%%%%%%%%%
%%%%%%%%%%%%%%%%%%%%%%%%%%%%%%%%%%%%%%%%%%%%%%%%%%%%%%%%%%%%%%%%%%

\end{document}